\documentclass[aps,prl,twocolumn,showpacs,floatfix,nofootinbib,preprintnumbers,superscriptaddress,amsmath,amssymb]{revtex4-1}

\usepackage{graphicx}
\usepackage{epsfig}
\usepackage{bm}
\usepackage{color}
\usepackage{float}
\usepackage{dcolumn}
\usepackage{multirow} 
\usepackage{hyperref}

\newcommand{\beq}{\begin{equation}}
\newcommand{\eeq}{\end{equation}}
\newcommand{\beqa}{\begin{eqnarray}}
\newcommand{\eeqa}{\end{eqnarray}}

\newcommand{\ChEFT}{$\chi$EFT}

\begin{document}

\title{Effects of three-nucleon forces and two-body currents on Gamow-Teller strengths}

\author{A.~Ekstr\"om} \affiliation{Department of Physics
   and Center of Mathematics for Applications, University of Oslo,
   N-0316 Oslo, Norway}

\author{G.~R.~Jansen} \affiliation{Physics Division, Oak Ridge National
Laboratory, Oak Ridge, TN 37831, USA}
\affiliation{Department of Physics and Astronomy, University of Tennessee,
Knoxville, TN 37996, USA}

\author{K.~A.~Wendt} \affiliation{Department
  of Physics and Astronomy, University of Tennessee, Knoxville, TN
  37996, USA} \affiliation{Physics Division, Oak Ridge National
  Laboratory, Oak Ridge, TN 37831, USA} 

\author{G.~Hagen} \affiliation{Physics Division, Oak Ridge National Laboratory,
Oak Ridge, TN 37831, USA} 
  \affiliation{Department of Physics and Astronomy, University of Tennessee,
  Knoxville, TN 37996, USA} 

\author{T.~Papenbrock} \affiliation{Department
  of Physics and Astronomy, University of Tennessee, Knoxville, TN
  37996, USA} \affiliation{Physics Division, Oak Ridge National
  Laboratory, Oak Ridge, TN 37831, USA} 

\author {S. Bacca}
\affiliation{TRIUMF, 4004 Wesbrook Mall, Vancouver, BC, V6T 2A3, Canada }

\author{B. Carlsson} \affiliation{Department of Fundamental Physics,
  Chalmers University of Technology, SE-412 96 G\"oteborg, Sweden}

\author{D. Gazit}
\affiliation{Racah Institute of Physics, Hebrew University, 91904, Jerusalem}

\begin{abstract}
  We optimize chiral interactions at next-to-next-to leading order to
  observables in two- and three-nucleon systems, and compute
  Gamow-Teller transitions in $^{14}$C and $^{22,24}$O using
  consistent two-body currents. We compute spectra of the daughter
  nuclei $^{14}$N and $^{22,24}$F via an isospin-breaking
  coupled-cluster technique, with several predictions. The two-body
  currents reduce the Ikeda sum rule, corresponding to a quenching
  factor $q^2 \approx 0.84-0.92$ of the axial-vector coupling.  The half
  life of $^{14}$C depends on the energy of the first excited $1^+$
  state, the three-nucleon force, and the two-body current.
\end{abstract}


\maketitle 

{\it Introduction.} -- $\beta$ decay is one of the most interesting
processes and most useful tools in nuclear physics. On the one hand,
searches for neutrino-less double-$\beta$ decay probe physics beyond
the standard model and basic properties of the neutrino, see
~\textcite{avignone2008} for a recent review. If neutrinoless
double-$\beta$ decay is observed, an accurate nuclear-physics matrix
element is needed to extract neutrino masses from the life time. On
the other hand, $\beta$ decay of rare isotopes populates states in
exotic nuclei and thereby serves as a spectroscopic
tool~\cite{winger2009,miernik2013}.  The theoretical calculation of
electroweak transition matrix elements in atomic nuclei is a
challenging task, because it requires an accurate description of the
structure of the mother and daughter nuclei, and an employment of a
transition operator that is consistent with the Hamiltonian.

For the transition operator, the focus is on the role of
meson-exchange currents~\cite{schiavilla2002} and two-body currents
(2BCs) from chiral effective field theory (\ChEFT).  Two-body currents
are related to three-nucleon forces (3NFs)
~\cite{fujita1957,vankolck1994} because the low energy constants
(LECs) of the latter constrain the former within \ChEFT.  Consistency
of Hamiltonians and currents is one of the hallmarks of an
EFT~\cite{epelbaum2009}, and 2BCs are applied in electromagnetic
processes of light nuclei,
see~\textcite{gazit2009,griesshammer2012,pastore2013}.  For weak
decays, only the calculation of triton $\beta$ decay
~\cite{gazit2009}, the related $\mu$ decay on $^3$He and the
deuteron~\cite{marcucci2012}, and proton-proton
fusion~\cite{marcucci2013}, exhibits the required consistency, while
the very recent calculation of the neutral-current response in
$^{12}$C employs phenomenological 3NFs together with chiral
2BCs~\cite{lovato2014}.  

The one-body operator $g_A\sum_{i=1}^A\sigma_i \tau_i^{\pm}$ induces
Gamow-Teller transitions. Here $g_A$ is the axial-vector coupling,
$\sigma$ denotes the spin, and $\tau^\pm$ changes the isospin.
Gamow-Teller strength functions~\cite{zegers2007,fujita2014} are of
particular interest also because of their astrophysical
relevance~\cite{langanke2003}. Charge-exchange measurements on
$^{90}$Zr and other medium mass nuclei have suggested that the total
strength for $\beta$ decay is quenched by a factor of $q^2 \approx
0.88-0.92$~\cite{sakai1999,yako2005,ichimura2006,sasano2009} when
compared to the Ikeda sum rule~\cite{ikeda1963}. Similarly,
shell-model calculations~\cite{chou1993,martinez1996} suggest that
$g_A$ needs to be quenched by a factor $q \approx 0.75$ to match
data.  It is not clear whether renormalizations (including 2BCs) of
the employed Gamow-Teller operator, missing correlations in the
nuclear wave functions, or model-space truncations are the cause of
this quenching.

Recent calculations~\cite{vaintraub2009,menendez2011,engel2014} show
that chiral 2BCs yield an effective quenching of $g_A$. However, the
Hamiltonians employed in these works are not consistent with the
currents (and they contain no 3NFs), and/or the 2BCs are approximated
by averaging the second nucleon over the Fermi sea of symmetric
nuclear matter. The recent studies~\cite{pervin2007,maris2011} of
electroweak transitions in light nuclei employ 3NFs but lack 2BCs.
This gives urgency for a calculation of weak decays that employs 3NFs
and consistent 2BCs.

In this Letter, we address the quenching of $g_A$ and employ 3NFs
together with consistent 2BCs for the computation of $\beta$ decays
and the Ikeda sum rule. We study the $\beta$ decays of $^{14}$C and
$^{22,24}$O with interactions and currents from \ChEFT~at
next-to-next-to leading order (NNLO) for cutoffs
$\Lambda_\chi=450,500,550$~MeV. For the states of the daughter nuclei,
we generalize a coupled-cluster technique and compute them as
isospin-breaking excitations of the mother nuclei.  We present
predictions and spin assignments for the exotic isotopes $^{22,24}$F,
and revisit the anomalously long half life of
$^{14}$C~\cite{holtjw2008,holtjw2009,maris2011}.

{\it Hamiltonian and model space.} -- The chiral nucleon-nucleon
($NN$) interactions are optimized to the proton-proton and the
proton-neutron scattering data for laboratory scattering energies
below 125~MeV, and to deuteron observables. The $\chi^2/\rm datum$
varies between 1.33 for $\Lambda_\chi=450$~MeV and 1.18 for
$\Lambda_\chi=550$~MeV. The $\chi^2$-optimization employs the
algorithm {\sc POUNDerS}~\cite{kortelainen2010}.  Table~\ref{tab1}
shows the parameters of the $NN$ interaction for the cutoff
$\Lambda_\chi=500$~MeV; the parameters for the other cutoffs are
supplementary material. The parameters displayed in Table~\ref{tab1}
are close to those of the chiral interaction NNLO$_{\rm
  opt}$~\cite{ekstrom2013}.

\begin{table}[hbt]
\begin{tabular}{|cc|cc|cc|} \hline \hline
LEC &  value & LEC &  value & LEC & value \\ \hline 
$c_1$       & -0.91940746 & $c_3$ &-3.88983848 & $c_4$ &   4.30736747 \\
$\tilde{C}^{pp}_{^1S_0}$  & -0.15136364  & $\tilde{C}^{np}_{^1S_0}$  & -0.15215263 & $\tilde{C}^{nn}_{^1S_0}$  & -0.15180482 \\ 
$C_{^1S_0}$ &   2.40431235 & $C_{^3S_1}$ &  0.92793712 & $\tilde{C}_{^3S_1}$ &-0.15848125 \\
$C_{^1P_1}$ &   0.41482908  & $C_{^3P_0}$ &   1.26578978 & $C_{^3P_1}$ & -0.77998484 \\
$C_{^3S_1-^3D_1}$ &  0.61855040  & $C_{^3P_2}$  & -0.67347042  &&\\ \hline \hline
\end{tabular}
\caption{ Pion-nucleon LECs $c_i$ and partial-wave contact LECs ($C$,
  $\tilde{C}$) for the chiral $NN$ interaction at NNLO using
  $\Lambda_\chi=500$ MeV and $\Lambda_{\rm
    SFR}=700$~MeV~\cite{epelbaum2004}. The $c_i$, $\tilde{C}_i$, and
  $C_i$ have units of GeV$^{-1}$, $10^4$ GeV$^{-2}$, and $10^4$
  GeV$^{-4}$, respectively.}
\label{tab1}
\end{table}

\begin{figure}[htb]
\includegraphics[width=0.48\textwidth]{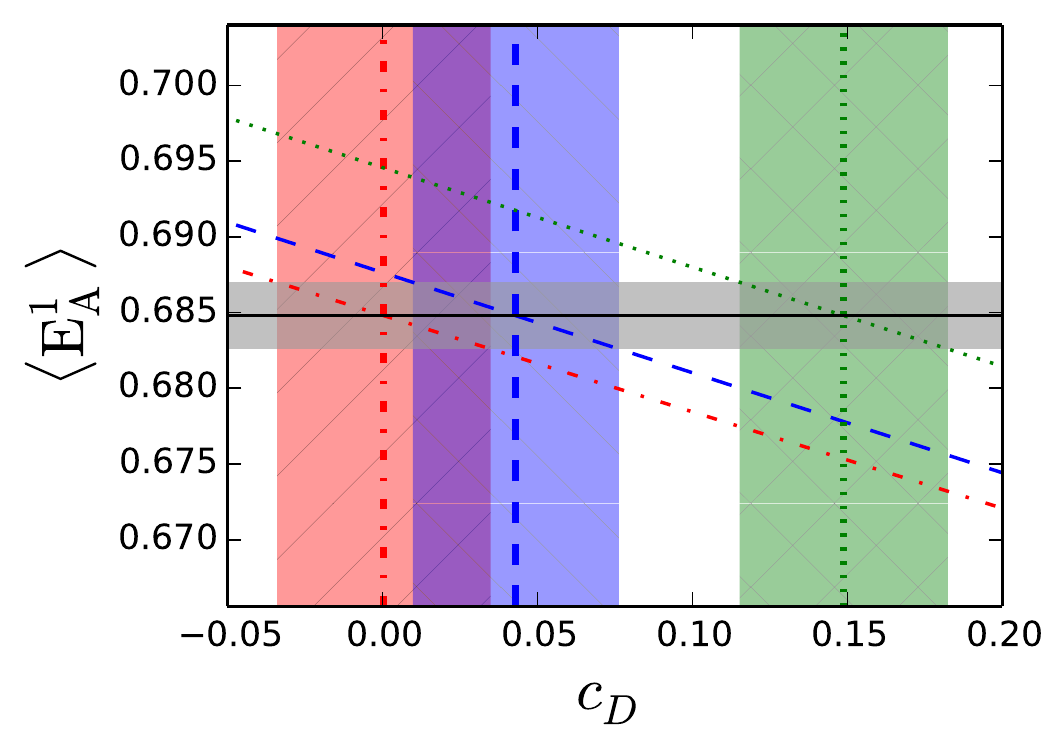}
\caption{(Color online) The quantity related to the triton half life
  $\langle E_1^A\rangle$ as a function $c_D$ 
  for chiral cutoffs $\Lambda_\chi=450,500,550$~MeV (red dashed-dotted,
  blue dashed, green dotted, respectively) with corresponding error
  bands. The different lines was determined by a
  fit of $c_D$ and $c_E$ to $A=3$ binding energies.}
  \label{error_bands}
\end{figure}

The 3NF is regularized with nonlocal cutoffs
\cite{epelbaum2002,hebeler2012} (to mitigate the convergence problems
documented by~\textcite{hagen2014} for local cutoffs). Following
\textcite{gazit2009}, we optimize the two LECs ($c_D$ and $c_E$) of
the 3NF to the ground-state energies of $A=3$ nuclei and the triton
lifetime.  Figure~\ref{error_bands} shows the reduced transition
matrix element $\langle E_1^A\rangle = \langle ^3{\rm He}\vert\vert
E_1^A\vert\vert ^3{\rm H}\rangle$ as a function of $c_D$. Here $E_1^A$
is the $J=1$ electric multipole of the weak axial vector current at
NNLO~\cite{gazit2009}. The leading-order (LO) contribution to $E_1^A$
is proportional to the one-body Gamow-Teller operator,
$E_1^A\vert_{\rm LO}=ig_A(6\pi)^{-1/2}\sum_{i=1}^A\sigma_i
\tau_i^{\pm}$. For the current we use the empirical value $g_A =
1.2695(29)$. The 2BCs enter at NNLO and depend on the LECs $c_D, c_3,
c_4$ of the chiral interaction~\cite{park2003,gazit2008}. The triton
half-life yields an empirical value for $\langle E_1^A\rangle_{\rm
  emp}$, which constrains $c_D$ and $c_E$. For the three different
chiral cutoffs $\Lambda_\chi=450,500,550$ the sets of $(c_D,c_E)$ that
reproduce the triton half-life and the $A=3$ binding energies are
$(0.0004,-0.4231), (0.0431,-0.5013), (0.1488,-0.7475)$, respectively.
The vertical bands in Fig.~\ref{error_bands} give the range of $c_D$
that reproduce $\langle E_1^A\rangle_{\rm emp}$ within the
experimental uncertainty.

We employ an $N=12$ model space consisting of $N+1$ oscillator shells
with frequency $\hbar\Omega=22$~MeV. The 3NFs use an energy
cutoff of $E_{\rm 3max}=N\hbar\Omega$, i.e. the sum of the excitation
energies of three nucleons does not exceed $E_{\rm 3max}$. 
We employ the intrinsic Hamiltonian
\begin{equation}
H= T-T_{\rm cm} +V_{NN} +V_{\rm 3NF}
\end{equation}
to mitigate any spurious center-of-mass
excitations~\cite{hagen2009a,jansen2012}. Here, $T$ and $T_{\rm cm}$
are the kinetic energy and the kinetic energy of the center-of-mass,
while $V_{NN}$ and $V_{\rm 3NF}$ are the chiral $NN$ interaction and
3NF, respectively.

We perform a Hartree-Fock (HF) calculation and compute the
normal-ordered Hamiltonian $H_{\rm N}$ with respect to the resulting
reference state $|{\rm HF}\rangle$.  We truncate $H_{\rm N}$ at the
normal-ordered two-body level, and note that 3NFs contribute to the
vacuum energy, and the normal-ordered one-body and two-body terms.
This approximation is accurate in light and medium-mass nuclei
\cite{hagen2007a,roth2012}.

{\it Formalism.} -- We compute the closed-subshell mother nuclei
$^{14}$C and $^{22,24}$O with the coupled-cluster
method~\cite{coester1958,coester1960,cizek1966,cizek1969,kuemmel1978,dean2004,bartlett2007,hagen2013c}.
The similarity-transformed Hamiltonian
\begin{equation}
\label{hamsim}
\overline{H}\equiv e^{-T}H_{\rm N}e^T
\end{equation}
employs the cluster amplitudes 
\begin{eqnarray}
T=\sum_{ia}t_i^a N^\dagger_a N_i +{1\over 4}\sum_{ijab}t_{ij}^{ab} N^\dagger_a N^\dagger_b N_j N_i 
\end{eqnarray}
that create 1-particle -- 1-hole (1p-1h) and 2-particle -- 2-hole
(2p-2h) excitations.  Here, $i,j$ denote occupied orbitals of the HF
reference while $a,b$ denote orbitals of the valence space. The
operators $N^\dagger_q$ and $N_q$ create and annihilate a nucleon in
orbital $q$, respectively. It is understood that the cluster
amplitudes $T$ do not change the number of protons and neutrons, i.e.
they conserve the $z$-component $T_z$ of isospin. We note that $|{\rm
  HF}\rangle$ is the right ground state of the non-Hermitian
Hamiltonian $\overline{H}$. Its left ground state is not $\langle {\rm
  HF}|$ but $\langle\Lambda| = \langle{\rm HF}\vert(1+\Lambda)$, with
$\Lambda$ being a linear combination of 1p-1h and 2p-2h de-excitation
operators~\cite{bartlett2007,hagen2013c}.

The daughter nuclei $^{14}$N and $^{22,24}$F are computed via a novel
generalization of the coupled-cluster equation-of-motion approach
\cite{stanton1993,gour2006,jansen2011}. We view the states of the
daughter nuclei as isospin-breaking excitations $|R\rangle\equiv
R|{\rm HF}\rangle$ of the coupled-cluster ground state, with
\begin{equation}
\label{R}
R\equiv \sum_{ia} r_i^a p^\dagger_a n_i + {1\over 4}\sum_{ijab} r_{ij}^{ab} p^\dagger_a N^\dagger_b N_j n_i \ .
\end{equation}
Here, $p^\dagger_q$ and $p_q$ ($n^\dagger_q$ and $n_q$) create and
annihilate a proton (neutron) in orbital $q$.  The combination
$N^\dagger_q N_s$ either involves neutrons $N^\dagger_q N_s =
n^\dagger_q n_s$ or protons $N^\dagger_q N_s = p^\dagger_q p_s$.  We
note that $R$ lowers the isospin component $T_z$ of the HF reference
by one unit and keeps the mass number unchanged. 

The states of the daughter nucleus result from solving the eigenvalue
problem $\overline{H}R_\alpha \vert {\rm HF}\rangle = \omega_\alpha
R_\alpha \vert {\rm HF}\rangle$. Here, $\omega_\alpha$ is the
excitation energy with respect to the HF reference, and $R_\alpha$
denotes a set of amplitudes
$R_\alpha=(r_i^a(\alpha),r_{ij}^{ab}(\alpha))$.  We recall that the
similarity-transformed Hamiltonian in Eq.~(\ref{hamsim}) is not
Hermitian.  Therefore, we also introduce the left-acting de-excitation
operator
\begin{equation}
\label{L}
L\equiv \sum_{ia} l^i_a n^\dagger_i p_a + {1\over 4}\sum_{ijab} l^{ij}_{ab} n^\dagger_i N^\dagger_j N_b p_a \ , 
\end{equation}
and solve the left eigenvalue problem $\langle {\rm HF}\vert L_\beta
\overline{H} = \omega_\beta \langle {\rm HF}\vert L_\beta$. The left
and right eigenvectors are bi-orthogonal, i.e. $\langle{\rm HF}|
L_\alpha R_\beta|{\rm HF}\rangle = \sum_{ia} l^i_a(\alpha)
r_i^a(\beta) +{1\over 4} \sum_{ijab} l^{ij}_{ab}
(\alpha)r_{ij}^{ab}(\beta)=\delta_{\alpha\beta}$.

The operators $R$ and $L$ in Eqs.~(\ref{R}) and (\ref{L}) excite
states in the daughter nucleus that results from $\beta^-$ decay. If
instead we were interested in $\beta^+$ decay, we would employ
$R^\dagger$ and $L^\dagger$, and solve the corresponding eigenvalue
problems.  Our approach allows us to compute excited states in the
daughter nucleus that are dominated by isospin-breaking 1p-1h
excitations of the closed-shell reference $|{\rm HF}\rangle$ (with
2p-2h excitations being smaller corrections).

\begin{figure}[htb]
\includegraphics[width=0.48\textwidth]{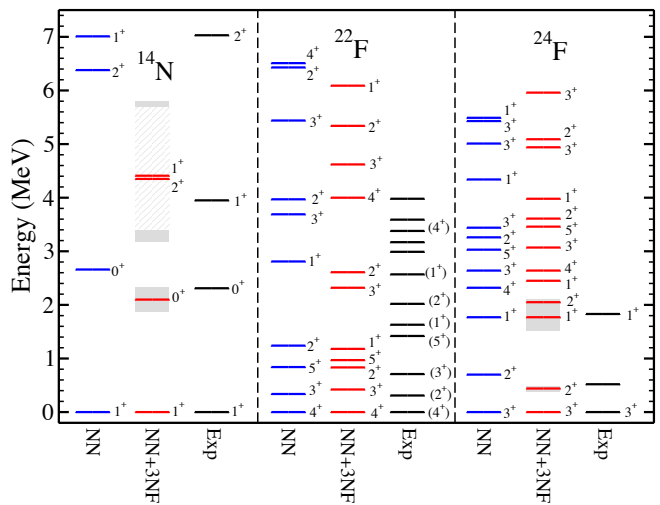}
\caption{(Color online) Spectra of the odd-odd daughter nuclei
  $^{14}$N and $^{22,24}$F resulting from the $NN$ interaction with
  chiral cutoff $\Lambda_\chi=500$~MeV (blue), the $NN$ interaction
  and 3NF at NNLO with chiral cutoff $\Lambda_\chi=500$~MeV (red),
  compared to experiment (black). Errorbars from variation of the
  chiral cutoff $\Lambda_\chi = 450$ to $550$~MeV are shown for the
  $0^+,2^+,1^+$, and the $2^+, 1^+$ excited states in $^{14}$N and
  $^{24}$F, respectively. The band with diagonal gray lines in
  $^{14}$N is for the $1^+$ excited state.}
  \label{spectra} 
\end{figure}

{\it Results.} -- The spectra for $^{14}$N and $^{22,24}$F are shown
in Fig.~\ref{spectra} for $\Lambda_\chi = 500$~MeV and compared to
data. Errorbars from variation of the chiral cutoff $\Lambda_\chi$ are
shown for selected states. The odd-odd daughter nuclei $^{14}$N and
$^{22,24}$F exhibit a higher level density than their mother
nuclei. Overall, 3NFs increase the level densities slightly and yield
a slightly improved comparison to experiment.  For the neutron-rich
isotopes of fluorine we make several predictions and spin assignments.
In these isotopes, our spectra compare also well to shell-model
calculations by \textcite{brown2006}. The ground state energies of the
mother nuclei (obtained at $N=12, \hbar\Omega=22$~MeV and
$\Lambda_\chi = 500$~MeV) are $-74.4$~MeV, $-104.6$~MeV, and
$-105.7$~MeV for $^{14}$C, and $^{22,24}$O, respectively. Thus, these
nuclei are significantly underbound compared to experiment. Our
calculations employ the same nucleon mass for protons and neutrons,
and we find the ground-state energies of the daughter nuclei are
$0.54$~MeV, $-2.62$~MeV, and $-6.55$~MeV with respect to their
corresponding mother nuclei, and in fair agreement with experiment.

Within the coupled-cluster framework we compute the total strengths
\begin{eqnarray*}
S_+ = \langle\Lambda| \overline{\hat{O}_{\rm GT}}\cdot\overline{\hat{O}_{\rm GT}^\dagger}|{\rm HF}\rangle \ , \ 
S_- = \langle\Lambda| \overline{\hat{O}_{\rm GT}^\dagger}\cdot\overline{\hat{O}_{\rm GT}}|{\rm HF}\rangle 
\end{eqnarray*}
for $\beta^\pm$ decays.  Here $\overline{\hat{O}_{\rm GT}}$ is the
similarity-transformed Gamow-Teller operator
\begin{equation}
\hat{O}_{\rm GT}\equiv \hat{O}_{\rm GT}^{(1)} +\hat{O}_{\rm GT}^{(2)}
\equiv g_A^{-1}\sqrt{3\pi} E_1^A\ .
\label{GTop}
\end{equation}
The one-body operator is $\hat{O}_{\rm GT}^{(1)} = g_A^{-1}\sqrt{3\pi}
E_1^A\vert_{\rm LO}$, and the two-body operator $\hat{O}_{\rm
  GT}^{(2)}$ is from the 2BC at NNLO~\cite{park2003,gazit2008}.

The Ikeda sum rule is $S_--S_+=3(N-Z)$ for $\hat{O}_{\rm GT}
=\hat{O}_{\rm GT}^{(1)}$. This identity served as a check of our
calculations. Our interest, of course, is in the contribution of the
2BC operator $\hat{O}_{\rm GT}^{(2)}$ to the total $\beta$ decay
strengths $S_\pm$. We considered two approximations of this two-body
operator. In the normal-ordered one-body approximation (NO1B), the
second fermion of the 2BC is summed over the occupied states of the HF
reference.  In the second approximation we add the leading order (LO)
contribution of the similarity transformed two-body operator,
$\overline{\hat{O}_{\rm GT}^{(2)}}\approx\hat{O}_{\rm GT}^{(2)}$ to
the NO1B contribution.  We will see below that this LO contribution is
a smaller correction to the NO1B contribution for the nuclei we study.

\begin{figure}[htb]
\includegraphics[width=0.48\textwidth]{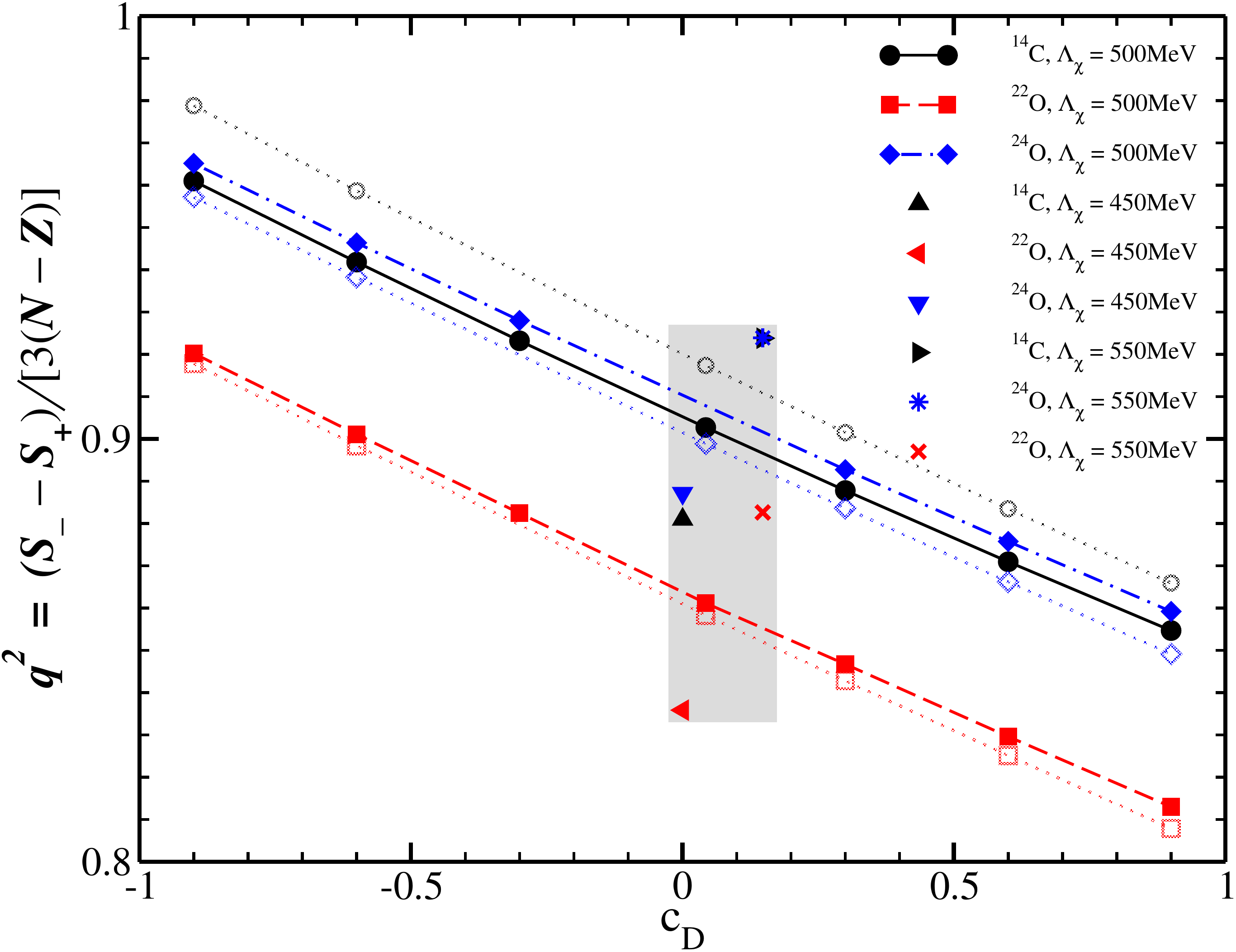}
\caption{(Color online) The quenching factor $q^2$ for $^{14}$C (black
  line), $^{22}$O (red dashed line), and $^{24}$O (blue dashed-dotted
  line) for different $c_D$ values. The calculations used $NN$ and 3NF
  with consistent 2BCs. The gray area marks the region of $c_D$ that
  yields the triton half life and shows the cutoff dependence. The
  dotted lines show the NO1B result.}
  \label{GT_sumrule}
\end{figure}

Figure~\ref{GT_sumrule} shows the quenching factor
$q^2=(S_--S_+)/[3(N-Z)]$ for $^{14}$C, and $^{22,24}$O. For the cutoff
$\Lambda_\chi=500$~MeV we vary $c_D$ between $-0.9$ and $0.9$ and fix
$c_E$ such that the binding energies of the $A=3$ nuclei are
reproduced. The ground-state energies and excited states in $^{14}$C
and $^{22,24}$F are insensitive to this variation. Thus, the
dependence of $(S_--S_+)/[3(N-Z)]$ on $c_D$ is due to 2BCs. The dotted
lines show the NO1B result. Thus, a major part of the quenching
results from the NO1B approximation. The sensitivity of our results to
the chiral cutoffs ($\Lambda_\chi=450,500,550$~MeV) is shown as the
gray band for values of $c_D$ and $c_E$ that reproduce the triton
half-life.  The quenching factor depends on the nucleus, with $q^2
\approx 0.84-0.92$ due to 2BCs for the studied nuclei.  We recall that
$q^2 \approx 0.88-0.92$, extracted from experiments on
$^{90}$Zr~\cite{yako2005,ichimura2006,sasano2009}, are within our
error band.  We also computed the low-lying strengths for $\beta^-$
decay, and found that only 70\% - 80\% of the total strength $S_\pm$
is exhausted below 10~MeV of excitation energy.

\begin{figure}[htb]
\includegraphics[width=0.48\textwidth]{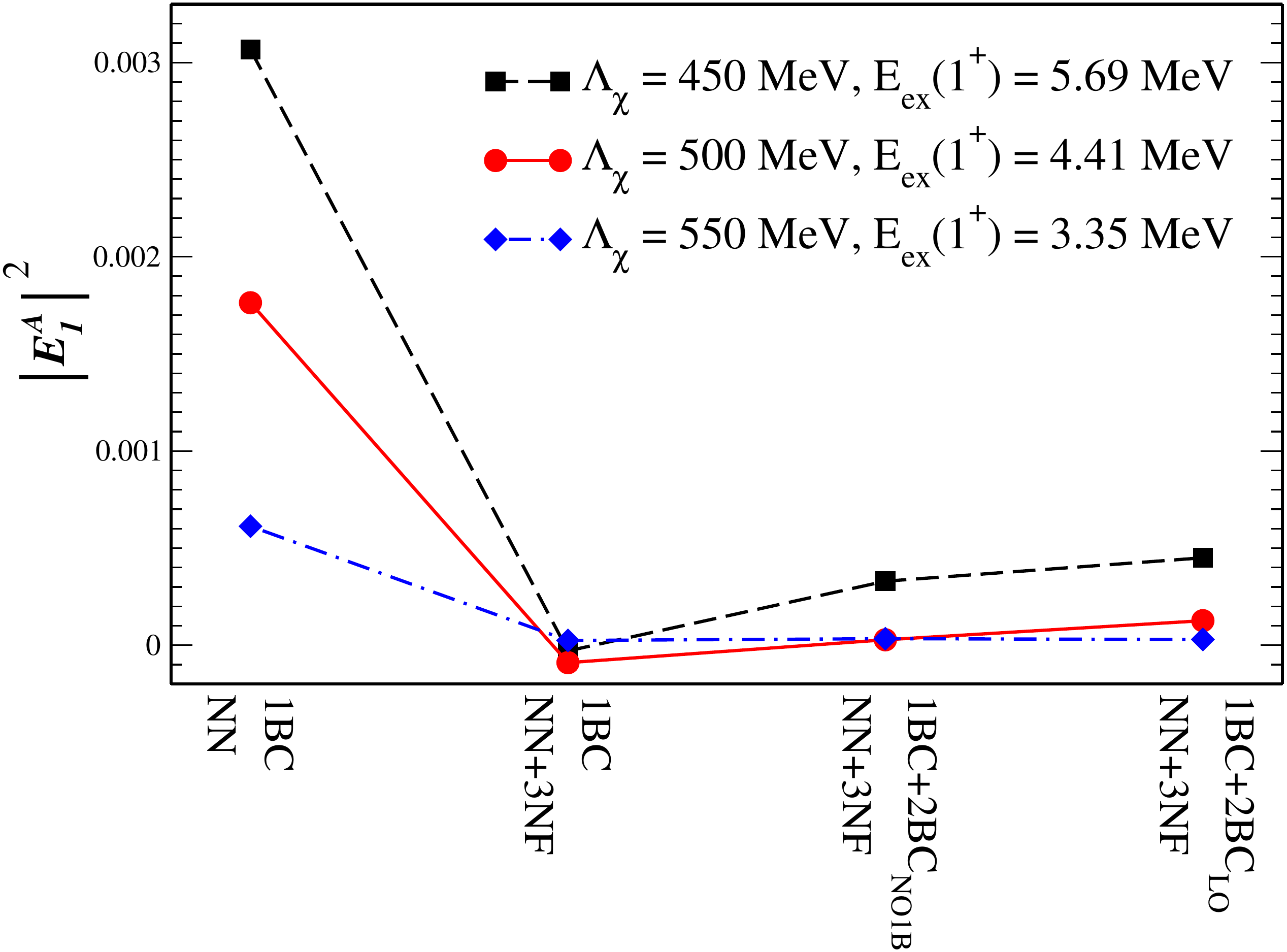}
\caption{(Color online) The squared transition matrix element for
  $\beta^-$ decay of $^{14}$C from increasingly sophisticated
  calculations (from left to right). NN, 1BC: $NN$ interactions and
  one-body currents (1BC) only.  NN + 3NF, 1BC: addition of 3NF.  NN +
  3NF, 1BC + 2BC$_{\rm NO1B}$: addition of 2BC in the NO1B
  approximation. NN + 3NF, 1BC + 2BC$_{\rm LO}$: addition of
  leading-order 2BC.}
\label{c14}
\end{figure}

Let us finally turn to the $\beta^-$ decay of $^{14}$C. The long half
life of this decay, about 5700~$a$, is used in carbon dating of
organic material. This half life is anomalously long in the sense that
it exceeds the half lives of neighboring $\beta$ unstable nuclei by
many orders of magnitude. Recently, several studies attributed the
long half life of $^{14}$C to
3NFs~\cite{holtjw2008,holtjw2009,maris2011}, while the experiment
points to a complicated strength function~\cite{negret2006}. What do
2BCs contribute to this picture? To address this question, we compute
the matrix element $\langle E_1^A\rangle\equiv \langle ^{14}{\rm
  N}|E_1^A|^{14}{\rm C}\rangle$ that governs the $\beta^-$ decay of
$^{14}$C to the ground state of $^{14}$N, with $c_D$ and $c_E$ from
the triton life time. Figure~\ref{c14} shows the various contributions
to the matrix element.  In agreement with
\textcite{holtjw2009,maris2011}, 3NFs reduce the matrix element
significantly in size, and our result is similar in magnitude as
reported by \textcite{maris2011}. However, 2BCs counter this reduction
to some extent, with the NO1B approximation and the LO approximation
both giving significant contributions. Our results for $\langle
E_1^A\rangle$ from 2BCs and 3NFs are between $5\times 10^{-3}$ and
$2\times 10^{-2}$. This is more than an order of magnitude larger than
the empirical value $\langle E_1^A\rangle_{\rm emp}\approx 6\times
10^{-4}$ extracted from the 5700~$a$ half life of $^{14}$C.

We also find that the matrix element $\langle E_1^A\rangle $ depends
on the energy of the first excited $1^+$ state in $^{14}$N. For the
three different cutoffs $\Lambda_\chi=450,500,550$~MeV this excited
$1^+$ state is at $5.69,4.41,3.35$~MeV, respectively (compared to
3.95~MeV from experiment). As the value of $\langle E_1^A\rangle$
decreases strongly with decreasing excitation energy, a correct
description of this state is important for the half-life in $^{14}$C.

{\it Summary.} -- We studied $\beta^-$ decays of $^{14}$C, and
$^{22,24}$O. Due to 2BCs we found a quenching factor $q^2 \approx
0.84-0.92$ from the difference in total $\beta$ decay strengths
$S_--S_+$ when compared to the Ikeda sum rule value $3(N-Z)$.  To
carry out this study, we optimized interactions from \ChEFT~at NNLO to
scattering observables for chiral cutoffs
$\Lambda_\chi=450,500,550$~MeV. We developed a novel coupled-cluster
technique for the computation of spectra in the daughter nuclei and
made several predictions and spin assignments in the exotic
neutron-rich isotopes of fluorine. We find that 3NFs increase the
level density in the daughter nuclei and thereby improve the
comparison to data. The anomalously long half life for the $\beta^-$
decay of $^{14}$C depends in a complicated way on 3NFs and 2BCs. While
the former increase the theoretical half life, the latter somewhat
counter this effect.  Taken together, the inclusion of 3NFs and 2BCs
yield an increase in the computed half life.

\begin{acknowledgments}
  We thank D. J. Dean, J. Engel, Y. Fujita, K. Hebeler, M.
  Hjorth-Jensen, M. Sasano, T. Uesaka, and A.  Signoracci for useful
  discussions. We also thank E. Epelbaum for providing us with
  nonlocal 3NF matrix elements.  This work was supported by the Office
  of Nuclear Physics, U.S.  Department of Energy (Oak Ridge National
  Laboratory), under DE-FG02-96ER40963 (University of Tennessee),
  DE-SC0008499 (NUCLEI SciDAC collaboration), NERRSC Grant No.\
  491045-2011, the Field Work Proposal ERKBP57 at Oak Ridge National
  Laboratory, and by the National Research Council and by the Nuclear
  Science and Engineering Research Council of Canada. DG's work is
  supported by BMBF ARCHES. Computer time was provided by the
  Innovative and Novel Computational Impact on Theory and Experiment
  (INCITE) program. This research used resources of the Oak Ridge
  Leadership Computing Facility located in the Oak Ridge National
  Laboratory, which is supported by the Office of Science of the
  Department of Energy under Contract No.  DE-AC05-00OR22725, and used
  computational resources of the National Center for Computational
  Sciences, the National Institute for Computational Sciences, and the
  Notur project in Norway.
\end{acknowledgments}

{\it Supplementary material.} -- The LECs for the NNLO interactions
with cutoffs $\Lambda=450,550$ MeV can be found in
Tables~\ref{tab2}-\ref{tab3}.
\begin{table}[hbt]
\begin{tabular}{|cc|cc|cc|} \hline \hline
LEC &  value & LEC &  value & LEC & value \\ \hline 
$c_1$       & -0.91029482 & $c_3$ & -3.88068766 & $c_4$ &  4.67092062 \\
$\tilde{C}^{pp}_{^1S_0}$  & -0.15203546  & $\tilde{C}^{np}_{^1S_0}$  & -0.15282740 & $\tilde{C}^{nn}_{^1S_0}$  & -0.15247258 \\ 
$C_{^1S_0}$ &  2.43109829   & $C_{^3S_1}$ & 0.98757436  & $\tilde{C}_{^3S_1}$ & -0.16953957 \\
$C_{^1P_1}$ &  0.46691821 & $C_{^3P_0}$ & 1.21516744 & $C_{^3P_1}$ & -0.85034985 \\
$C_{^3S_1-^3D_1}$ & 0.68142133  & $C_{^3P_2}$  & -0.67318268  &&\\ \hline \hline
\end{tabular}
\caption{ Pion-nucleon LECs $c_i$ and partial-wave contact LECs ($C$,
  $\tilde{C}$) for the chiral $NN$ interaction at NNLO using
  $\Lambda_\chi=450$ MeV and $\Lambda_{\rm
    SFR}=700$~MeV~\cite{epelbaum2004}. The $c_i$, $\tilde{C}_i$, and
  $C_i$ have units of GeV$^{-1}$, $10^4$ GeV$^{-2}$, and $10^4$
  GeV$^{-4}$, respectively.}
\label{tab2}
\end{table}

\begin{table}[hbt]
\begin{tabular}{|cc|cc|cc|} \hline \hline
LEC &  value & LEC &  value & LEC & value \\ \hline 
$c_1$       & -0.90630268 & $c_3$ & -3.89738533 & $c_4$ &   3.90628243 \\
$\tilde{C}^{pp}_{^1S_0}$  & -0.15067278  & $\tilde{C}^{np}_{^1S_0}$  & -0.15162371 & $\tilde{C}^{nn}_{^1S_0}$  & -0.15121579 \\ 
$C_{^1S_0}$ & 2.38965389   & $C_{^3S_1}$ & 0.83899578 & $\tilde{C}_{^3S_1}$ & -0.14677863 \\
$C_{^1P_1}$ &  0.38612051  & $C_{^3P_0}$ &  1.32532984 & $C_{^3P_1}$ &  -0.68424744 \\
$C_{^3S_1-^3D_1}$ & 0.56266120  & $C_{^3P_2}$  &  -0.67444090  &&\\ \hline \hline
\end{tabular}
\caption{ Pion-nucleon LECs $c_i$ and partial-wave contact LECs ($C$,
  $\tilde{C}$) for the chiral $NN$ interaction at NNLO using
  $\Lambda_\chi=550$ MeV and $\Lambda_{\rm
    SFR}=700$~MeV~\cite{epelbaum2004}. The $c_i$, $\tilde{C}_i$, and
  $C_i$ have units of GeV$^{-1}$, $10^4$ GeV$^{-2}$, and $10^4$
  GeV$^{-4}$, respectively.}
\label{tab3}
\end{table}

\bibliographystyle{apsrev}

\end{document}